# Performance Evaluation of Wavelength Routed Optical Network with Wavelength Conversion

Vitthal J. Gond and Aditya Goel

**Abstract –** The rapid development of telecommunication networks is driven by user demands for new applications and advances in technologies. The explosive growth of the internet traffic is due to its use for collecting the information, communication, multimedia application, entertainment, etc. These applications are imposing a tremendous demand for bandwidth capacity on telecommunication network. The introduction of fiber optics had proved to meet the huge demand of bandwidth. These requirement can be meet by all optical network which is capable of transmitting enormous data at very high speed, around 50 Tera bits per seconds (Tbps) A wavelength conversion technique is addressed in this paper to reduced the blocking probability in wavelength routed networks. It is seen that the blocking probability of traffic requests decreases as the wavelength conversion factor increases. We explode the possibility for network with different size with variation in wavelength per link. In this work the evaluation of wavelength routed optical network with varying number of wavelength converters, different traffic types are carried out and results are shown that the blocking probability is minimum with 50% to 60% wavelength convertible nodes. Wavelength convertible nodes more than 60% are not showing much effect on reduction in blocking probability rather it results in increase in overall cost of network.

——————————— ◆ ———————————

## 1. INTRODUCTION

Optical WDM networks meet the tremendous demand of the new world, as it possesses huge bandwidth [1,2,5]. The maximum rate at which an end-user can access the network is limited by electronic speed (to a few Gb/s), the key in designing optical communication networks in order to exploit the fiber's huge bandwidth is to introduce concurrency among multiple user. In an optical communication network, WDM network may provide this concurrency. In Wavelength Routed WDM networks wavelength conversion technique is used to increase the fiber link utilization and reducing the network blocking probability. But wavelength converters are very expensive so our objective is to achieve a good performance of network that means reduces blocking probability, improve link utilization at a reduced cost. In this work the effects of varying number of wavelength converters, different traffic types on fiber link utilization and network blocking probability are analyzed.

## 2. OPTICAL WDM NETWORK

An Optical WDM Network is consists of wavelength routing nodes interconnected by point-to-point optical fiber links in an arbitrary topology. A Routing and Wavelength Assignment (RWA) algorithm selects a good route and wavelength to satisfy a connection request so as to improve the network performance. Different RWA algorithms are studied in[6]. Basically WDM network architectures are of two type broadcast-and-select networks, wavelength routed networks. In broadcast-and-select networks different nodes transmit messages on different wavelengths simultaneously. The star coupler combines all these messages and then broadcasts the combined message to all the nodes. A node selects a desired wavelength to receive the desired message by tuning its receiver to that wavelength. The advantage of broadcast-and-select networks is their simplicity. However, they have limitations such as large number of wavelength are required, there is no wavelength reuse in the network, they cannot span long distances since the transmitted power is split among various nodes and each node receives only a small fraction of the transmitted power, which becomes smaller as the number of nodes increases. Where as wavelength routed WDM networks have the potential to avoid the above problems of broadcast-and-select networks.

A wavelength routed network consists of wavelength cross connects, interconnected by point-to-point fiber links in an arbitrary topology. In a wavelength routed network, a message is sent from one node to another node using a wavelength continuous route called a light path, without requiring Optical–Electronic–Optical (OEO) conversion thus reduces delay due to OEO conversion and buffering. A light path is an all-optical communication path between two nodes, established by allocating the same wavelength throughout the route of the transmitted data. The requirement that the same wavelength must be used on all the links along the selected route is known as the wavelength continuity constraint. Two light paths cannot be assigned the same wavelength on any fiber. This requirement is known as distinct wavelength assignment constraint. However, two light paths can use the same wavelength if they use disjoint sets of links. This property

————————————————

- *V. J. Gond is with the Department of Electronics and Communication,*
  *Maulana Azad National Institute of Technology, Bhopal-462051, India.*
- *A. Goel is with the Department of Electronics and Communication,*
  *Maulana Azad National Institute of Technology, Bhopal-462051, India.*

is known as wavelength reuse. Wavelength continuity constrains results in inefficient utilization of wavelength channel, bandwidth loss caused by the wavelength continuity constraint can be overcome by using a wavelength converters, it also helps in increasing the number of light paths established while employing a limited number of wavelengths. Thus Wavelength reuse in wavelength routed networks makes them more scalable. Another important characteristic which enables wavelength routed networks to span long distances is that the transmitted power invested in the light path is not split to irrelevant destinations. In a wavelength routed optical WDM network, each wavelength can be routed through the optical network at the optical switching nodes, removing the need for optical-electronic conversion and electronic routing. Data can remain in optical domain without requiring costly high-speed electronic equipment and eliminate the bottleneck due to O-E-O conversion at intermediate router nodes. However, to establish and operate a WDM network several issues play important role towards the efficiency and performance of the WDM networks. Some of these issues include routing and wavelength assignment, wavelength continuity constraint, reconfiguration and survivability of virtual topology (optical layer). In wavelength routed WDM networks, a connection is realized by a light path, in order to establish a connection between a source-destination node pair, a wavelength continuous route needs to be found between the node pair. The total number of bidirectional routes, R, in the network is given by

$$R = N*(N-1) \quad \text{------------} \quad (1)$$

Where as for unidirectional routes, R in the network is given as

$$R = N*(N-1)/2 \quad \text{------------} \quad (2)$$

As all the routes are equally likely in uniform traffic distribution, the total load on the network, L is uniformly distributed on all the routes. Therefore the load on any route r, Lr is given by

$$L_r = \frac{L}{R} \quad \text{-------} \quad (3)$$

An algorithm used for selecting routes and wavelengths to establish light paths is known as a routing and wavelength assignment (RWA) algorithm [3,6]. Routing and Wavelength Assignment algorithm may be static or dynamic [3]. A good wavelength assignment algorithm in a WDM network can result in improved performance. The route is chosen based on some criterion such as the hop length, and the wavelength is chosen based on some criterion such as wavelength usage factor in the entire network. Different RWA algorithms have been proposed in the literature[6] to choose the best pair of route and wavelength. They differs in their policies for selecting route and wavelength, different RWA algorithms are: Fixed Routing, Fixed Alternate Routing and Exhaust Routing. Similarly different wavelength assignment algorithms are Fixed Order, Maximum Used, List Used and Random Selection of wavelength. A wavelength converter is an optical device which is capable of shifting one wavelength to another wavelength. The capability of a wavelength converter is characterized by the degree of conversion, a converter is said to have full degree of wavelength conversion if it is capable of shifting any wavelength to any other wavelength. Otherwise, it is said to have partial or limited degree of wavelength conversion.

Given a physical topology, number of wavelengths, and resources constraints and the long-term average traffic flow between node pairs, the problem is to place optimal wavelength converters so as to optimize a certain metric such as blocking probability, link utilization, network congestion or message delay. The performance optimization metrics to be considered while designing a wavelength routed WDM network with wavelength converters are minimizing congestion in the network, maximizing single-hop traffic flow, minimizing message delay. The wavelength continuity constraint imposed by WDM networks results in inefficient utilization of wavelength channels. A request may have to be rejected even though a route is available because of not availability of the same wavelength on all the links along the route. Due to this inefficient utilization of wavelength channels, more connections are blocked. The wavelength continuity constraint can be relaxed at a node by placing optical wavelength converters at the nodes. A wavelength converter helps to improve network performance by relaxing the network continuity constraint. As converters are very expensive, it is not economically feasible to place converters at all nodes, where as determining the nodes where converters can be placed in order to optimize network performance is an important problem known as converter placement problem.

This problem can be stated as follows: Given the number of Node 'N', we selects 'K' nodes that are capable of converting wavelength, where K ≤ N and placing 'K' nodes such that this placement yields the best performance. Some of the factors that are to be considered while doing optimal converter placement are a node with heavy transit traffic could be a possible candidate for converter placement. However, if it does very little mixing (or switching) of traffic, it may not be a good choice for an optimal solution. A node which has less transit traffic is not a good choice.

Types of traffics we used in our simulation are Constant Bit Rate (CBR) and exponential traffic. CBR traffic generator generates traffic according to a deterministic rate and packets are of constant size. Exponential traffic generator generates traffic according to an Exponential On/Off distribution. Packets are sent at a fixed rate during on periods, and no packets are sent during off periods. Both on and off periods are taken from an exponential distribution. The effects of different data rates in a WDM network are evaluated and results are compared in this paper. Blocking probability is an important performance parameter, it is the probability of blocking a request for a connection. When a new request for a connection comes to a node it searches for the available resources free wavelength, free route and if the appropriate resources are not free then it rejects the request for the connection that is it blocks the request. To minimize the blocking of a request in a WDM network there should be sufficient number of wavelengths, sufficient number of routes, an optimum number of wavelength converters placed at appropriate nodes and sufficient conversion capability of wavelength converter.

Link utilization is the measure of utilized capacity of an optical fiber link in a WDM networks. Our objective is to achieve maximum link utilization. Link utilization is generally limited because of other factors like traffic rate, data rate, number of nodes, number of wavelengths etc.

## 3. SIMULATION

In the simulations, the RWA algorithm with fixed-alternate shortest path routing and first-fit wavelength assignment is analyzed. Sparse wavelength conversion scheme is used to model the wavelength conversion capability of a network. Each simulation has been evaluated with both Exponential and CBR session traffic. Randomly generated topologies and session traffic pairs to create a diverse set of scenarios are used in the simulations. This random topology presents an imitation of actual Wide-Area-Network (WAN) whose topology and traffic rate is not known a priory. The topologies that were used had 25, 50 75 and 100 nodes, with each link having 16, 32, 48 and 64 wavelengths. The results for 100 nodes network are discussed here.

## 4. RESULTS AND DISCUSSION

The results of performance measurements for 100-node WDM networks, which have 16, 32, 48 and 64 wavelengths on each multi-wavelength link respectively incorporated with Exponential and CBR traffic to observe the effects of wavelength conversion factor, effect of traffic load and effect of data rates are shown. The topology generator generated the physical topology with a connectivity probability of 0.03. Connection probabilities of 0.03 results in a partial mesh topology. When the connection probability is increased the topology moves toward full mesh topology. Since the number of links required in full mesh topology for 100 nodes is too large (approximately 10000) and thus not feasible in real network, therefore the connection probability was set at 0.03. In the simulations the traffic load is kept constant at 0.4 Erlangs so that performance of a WDM network can be studied without traffic overload. Figure 1 (a-d) shows the effects of wavelength conversion factor on blocking probability for various wavelength numbers. It is seen that the blocking probability of traffic requests decreases as the wavelength conversion factor increases. The rate of decreasing blocking probability is more in case of Constant Bit Rate (CBR) traffic, than the Exponential traffic, as shown in Figure 1 (a-d). At wavelengths 32 and 64 with exponential traffic, the blocking probability is almost same up to wavelength conversion factor 0.4 then there is a decrement in blocking probability up to wavelength conversion factor 0.5 and again almost constant curve. Thus with 32 and 64 wavelengths the number of nodes having wavelength converters should be 50 %, it is not useful to use more than 50 % wavelength convertible nodes because the performance is almost equal and wavelength convertible nodes are very expensive. It is observed that in case of 16 and 48 wavelengths, the wavelength conversion factor should be 0.7, thus 70 % wavelength convertible nodes are required to give good performance. With CBR traffic wavelengths 32 and 64 have least blocking probability.

Figure 2 (a-d) shows the effects of wavelength conversion factor on link utilization for various wavelength numbers with CBR traffic and exponential traffic. It is seen that the link utilization is better with CBR traffic (approx. up to 25-26 %) than the exponential traffic (approx. up to 19 %). As the number of wavelengths increases from 16 to 64 the utilization of the link decreases.

## 5. CONCLUSION

In the simulations It is seen that the blocking probability of traffic requests decreases as the wavelength conversion factor increases. The rate of decreasing blocking probability is more in case of Constant Bit Rate traffic than the Exponential traffic. With 32 and 64 wavelengths the number of nodes having wavelength converters should be 50 %. It is not useful to use more than 50 % wavelength convertible nodes because the performance is almost same. With CBR traffic wavelengths 32 and 64 have least blocking probability with a wavelength conversion factor of 0.5. Thus from the blocking probability point of view 32 and 64 wavelengths are proposed to be utilized in the network.

**Figure 1 (a) - (d) Effects of Wavelength Conversion Factor on Blocking Probability For different wavelengths a)W=16 b)W=32 c)W=48 d)W=64, Nodes=100, Connection Probability = 0.03, Traffic Load=0.4 E**

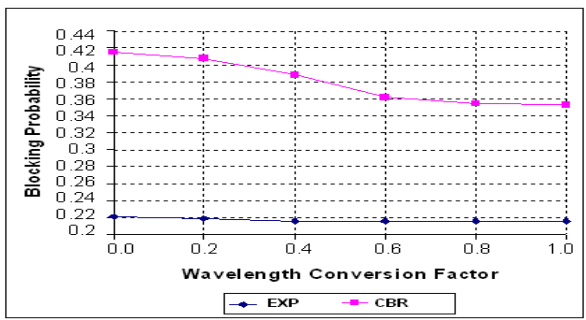

**Figure 1.a, W= 16**

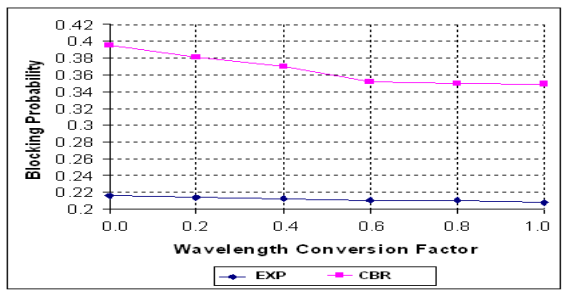

**Figure 1.b, W = 32**

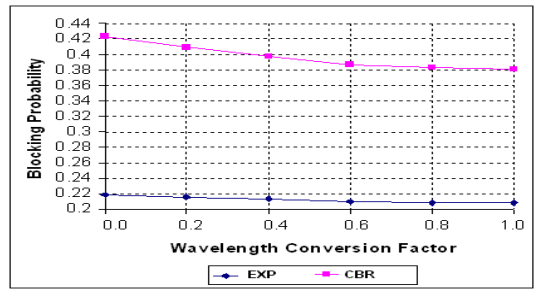

**Figure 1.c , W = 48**

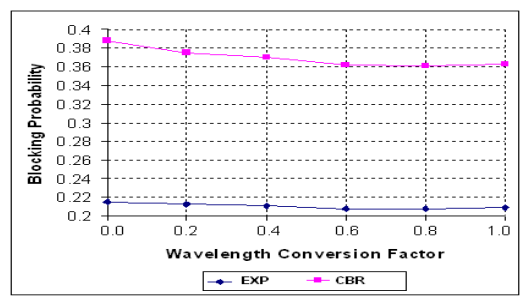

**Figure 1.d , W = 64**

**Figure 2(a)-(d) Effects of Wavelength Conversion Factor on Link Utilization for different wavelengths a)W=16 b)W=32 c) W=48 d)W=64, Nodes=100, Connection Probability = 0.03, Traffic Load=0.4 E**

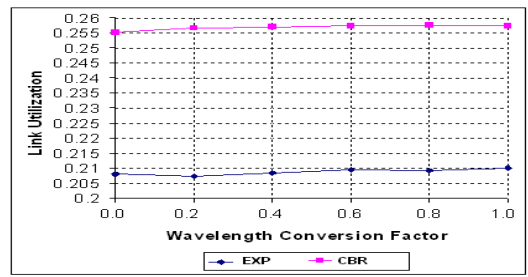

**Figure 2.a, W = 16**

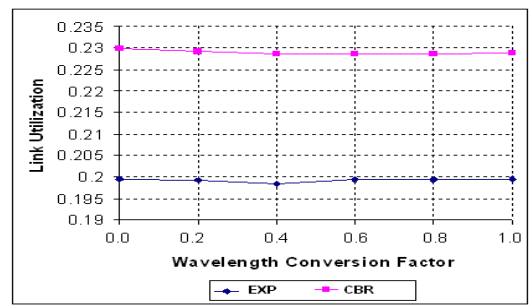

**Figure 2.b , W = 32**

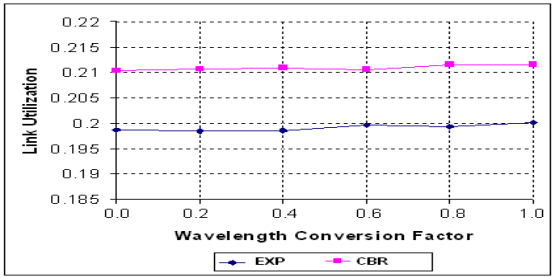

**Figure 2.c , W = 48**

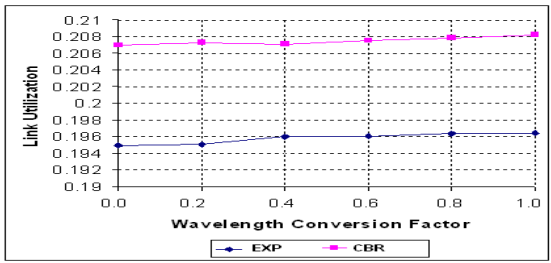

**Figure 2.d , W = 64**